\documentclass[12pt]{article}
\usepackage{graphicx}
\usepackage{epstopdf}
\usepackage[body={17.5cm, 22cm},right=2cm]{geometry}
\usepackage{amssymb}
\usepackage{amsmath}
\usepackage{bm}

\newcommand\ee{\end{equation}}
\newcommand\be{\begin{equation}}
\newcommand\eea{\end{eqnarray}}
\newcommand\bea{\begin{eqnarray}}

\def\beq{\begin{equation}}
\def\eeq{\end{equation}}

\newcommand\earr{\end{array}}
\newcommand\barr{\begin{array}}

\begin{document}
\setcounter{page}{0}
\thispagestyle{empty}

\begin{titlepage}
\begin{flushright}
 SLAC-PUB-14663
\end{flushright}

~\vspace{1cm}
\begin{center}

{\LARGE \bf Universality of the Volume Bound\\[.4cm]
in Slow-Roll Eternal Inflation}

\vspace{1.2cm}

{\large \bf
Sergei Dubovsky$^{a}$,
Leonardo Senatore$^{b,c}$, 
Giovanni Villadoro$^{d}$}
\\
\vspace{.6cm}
{\small {\sl $^a$ Center for Cosmology and Particle Physics, Department of Physics,\\
New York University, New York, NY, 10003, USA}}

\vspace{.3cm}
{\normalsize { \sl $^{b}$ Stanford Institute for Theoretical Physics,\\
Stanford University, Stanford, CA 94305 USA}}

\vspace{.3cm}
{\normalsize { \sl $^{c}$ Kavli Institute for Particle Astrophysics and Cosmology,\\
Stanford University and SLAC, Stanford, CA 94305 USA}}

\vspace{.3cm}
{\normalsize { \sl $^{d}$ SLAC, Stanford University \\
2575 Sand Hill Rd., Menlo Park, CA 94025 USA}}

\end{center}
\vspace{.8cm}
\begin{abstract}
It has recently been shown that in single field slow-roll inflation the total volume cannot grow by a factor larger than $e^{S_{dS}/2}$ without becoming infinite. 
The bound is saturated exactly at the phase transition to eternal inflation where the probability to 
produce infinite volume becomes non zero. 
We show that the bound holds sharply also in any space-time dimensions, 
when arbitrary higher-dimensional operators are included and in the multi-field inflationary case. 
The relation with the entropy of de~Sitter and the universality of the bound strengthen the case for a deeper holographic interpretation. As a spin-off we provide the formalism to compute the probability distribution of the volume after inflation for generic multi-field models, which might help to address questions about the population of vacua of the landscape during slow-roll inflation.
\end{abstract}

\end{titlepage}



\section{Introduction and Conclusions}

The future evolution of our Universe appears to be dominated by a phase of accelerated expansion~\cite{Riess:1998cb,Komatsu:2010fb}. 
The data from the cosmic microwave background strongly suggest that also in the distant past our Universe  experienced a phase
of slow-roll inflation~\cite{Guth:1980zm,Linde:1981mu,Albrecht:1982wi}. 
Furthermore, independently from the actual observations, the study of the physics in a accelerating universe is interesting in its own, as still several theoretical aspects are poorly understood. 

A particularly interesting set-up is represented by slow-roll eternal inflation~\cite{Vilenkin:1983xq,Linde:1986fd,Linde:1986fc}. In this case the scalar potential is so flat
that quantum fluctuations dominate over the classical rolling of the scalar field. 
In this limit the scalar field becomes free, so despite
quantum fluctuations dominate over the classical evolution the system becomes exactly solvable~\cite{Creminelli:2008es}.
In this case there is a finite probability to generate inflaton trajectories going uphill the
scalar potential, which can make inflation last forever. Quantum fluctuations are able to completely change the future causal structure of space-time, similarly to what happens in false vacuum eternal inflation~\cite{Guth:1982pn,Sekino:2010vc} or in the presence of a black-hole~\cite{Hawking:1974sw}.  Indeed, these are the only three known solutions of general relativity where quantum effects induce such a spectacular behaviour. In particular, in the eternal inflation case,
the result is even more dramatic because the background geometry becomes completely stochastic.

The dependence of the future causal structure of space-time on the dynamics
of the theory is a delicate issue already  at the semiclassical level. 
The problem is even more acute in full quantum gravity, where local observables are not well defined and its proper formulation in known cases (such as the string theory
$S$-matrix and $AdS$/CFT) strongly relies on well defined asymptotic boundaries.
To make the problem even more involved, unitarity, and in this particular context, holography also seem to suggest that 
space-time regions causally disconnected by an horizon are redundant, complementary.

These problems are particularly relevant in the framework of the landscape. 
If, as suggested by string theory, quantum gravity possesses a landscape of (meta-stable) vacua, our Universe may be doomed to deal with eternal inflation.
While vacuum tunnelling from meta-stable de~Sitter vacua seems the natural mechanism for generating eternal inflation in the landscape, slow-roll eternal
inflation represents a unique framework to reliably study the onset of the transition between the non-eternal and the eternal inflating regime.
In these type of models indeed, it exists a controllable parameter (the flatness of the potential) that smoothly interpolates between the two phases. It basically 
allows the study of the de~Sitter phase with a tunable parameter. 

In this context a bound was found in \cite{ArkaniHamed:2007ky} for any model of inflation in the non-eternal phase: the number of $e$-folding $N$ is always bounded by the de~Sitter entropy $S_{dS}$ at the end of inflation. The relation with the entropy seems to suggest a connection with the holographic bounds.
It is indeed in close analogy with the bound on the validity of the effective field theory description of the Hawking radiation quanta in the black-hole evaporation process:
after a time of order the black-hole entropy in units of the curvature scale the EFT predictions for the Hawking entropy start deviating by order one from what expected from the unitary evolution---information starts coming out and complementarity emerges. 
A similar time-scale shows up in slow-roll inflation exactly when the eternal inflation phase is approached.

It was found later in \cite{Creminelli:2008es} that in single field slow-roll inflation the phase transition to eternal inflation is sharp: at the critical value of the parameter
\begin{equation}
\Omega\equiv\frac{2\pi^2}{3} \frac{{\dot\phi}^2}{H^4}=1\,,\nonumber
\end{equation}
a finite probability of creating an infinite volume after inflation develops. This suggested the existence of a sharp bound also for the number of $e$-foldings.
Because of quantum fluctuation however, the number of $e$-foldings is not a well defined object, as it can fluctuate from point to point in space. 
The invariant quantity is the total volume of the Universe at the reheating surface, which matches $e^{3N}$ in the absence of quantum fluctuations. 
Quantum fluctuation makes this quantity stochastic too. A probability density for the volume $V$ can nonetheless be defined and actually computed. 
This was accomplished in \cite{Dubovsky:2008rf} where a sharp formulation for the bound was also found, namely:

\emph{The probability of producing a finite volume $V$ larger than $e^{S_{dS}/2}$ vanishes up to non-perturbative quantum gravity effects.}

Notice that the bound applies not only in the non-eternal regime ($\Omega>1$), where the volume is always finite, but also in the eternal regime, where there is still a non-zero probability to have a finite volume.

The existence of a sharp bound and the connection with the entropy clearly cries for an holographic interpretation. Is this bound really set by complementarity? 
In that case is there a meaning of the factor $\frac12$ appearing in the bound? We are not able to give a definitive answer to these questions yet.
It seems however that if the answers to these questions are positive then the bound should be universal, including the factor $\frac12$, as, for example, in the Bekenstein-Hawking formula. 

In this paper we test the universality of the bound against three different generalizations of the models studied in 
\cite{Dubovsky:2008rf} by changing the number of space-time dimensions (from 4 to $D$), by taking into account
 the effects of higher-dimensional operators in the action, and by considering more inflaton fields.

In the first test the dependence on the number of dimensions appears in a non-trivial way in a number of quantities entering
the calculation of the bound, such as the relation between the entropy and the horizon area, the Friedmann equations, the quantum
fluctuations of a inflaton field in de~Sitter space, etc. We will show that, surprisingly, when the bound is written in terms of the total volume
and the $dS_D$ entropy the dependence on the number of dimensions cancel out leaving the bound unchanged, with the same factor
as in four dimensions.

The second test we performed is with respect to higher derivative terms in the action. These terms not only
change the equations of motion for the metric and the inflaton, thus the Friedmann equations, but also the expression
for the entropy, which is not given anymore by just the horizon area in Planck units, and the size of the quantum fluctuations
of the inflaton. Remarkably we find that, independently of the corrections considered in the action, the bound is not affected,
included the factor $\frac12$, which stays universal.

The third test requires more effort. We have to extend the formalism developed in \cite{Dubovsky:2008rf} to the case of multifield
inflation. In particular the possibility to have extended regions in field space where inflation can end triggers difficulties both at the technical
and the conceptual level. We will show that the main formula that in the single-field case was a simple Laplace transform of the solution of a non-linear differential equation maps in this case into a functional integral transform of the solution of a non-linear partial differential equation. 
These more involved formulae allow us to calculate not only the probability distribution of the total volume after inflation, but the multi-variate probability distribution
of each different kind of volume associated with the different reheating points in field space.
The possibility of exiting inflation in different places in field space, where in particular the Hubble scale and thus the entropy are different, 
forces us to generalize the definition of the bound too. 
A conservative assumption is to assume that the total volume be bounded by the largest possible entropy
on the reheating surface. We believe however that a stronger version of the bound actually holds.

\emph{In all the realizations with finite total volume, the probability of producing a particular volume with values 
of the inflaton field within a given region ${\cal I}$ of the reheating surface and larger than the largest value of 
$e^{{S_{dS}}/2}$ on the same region ${\cal I}$ vanishes up to non-perturbative quantum gravity effects.}

In particular given any reheating region in field space the volume produced after inflation with that kind of inflaton values
will be bounded by the corresponding de~Sitter entropy.

Because of the complexity of the formulae involved we are not able to test the validity of the bound in the most general case,
however we will discuss the two simplest examples where we are able to take the calculation till the end. The two examples correspond to
two-field inflationary models with a constant slope of the potential and where the reheating region is a straight line normal or at angle
with respect to the slope direction.  We will be able to calculate the average volume distribution as a function of the reheating point
and show that the bound is indeed satisfied. Interestingly the factor $\frac{1}{2}$ is still universal and the bound is saturated only
when the slope is actually orthogonal to the reheating surface. This configuration correspond to single field inflation with a spectator field.
This result suggests that the presence of extra fields only makes the bounds stronger, by drifting the inflaton trajectories towards region
of higher entropy where the bound is less constraining.

It would be nice to test the bound also for more complex field configuration, but we are limited by our ability to
solve the corresponding equations.
Another interesting effect that we have neglected so far is the inclusion of slow-roll corrections.
While they are expected to be very small, after all in the eternal inflation limit the slow-roll parameter is tiny,
when considering inflaton trajectories that start saturating the bound, the integrated slow-roll effect may be non-negligible,
the tiny slow-roll parameter being compensated by the large field excursion. Arguments were given in \cite{Dubovsky:2008rf} why this effect should not alter the nature of the bound, still a dedicated study would be worthwhile.

Finally we would like to mention a suggestive relation between our bound and some 
thermodynamic relations recently found for systems out of equilibrium (see e.g.~\cite{Jarzynski1997}).
The variation of the entropy in a closed thermodynamical transformation can be thought of as a measure of the irreversibility of the process. If we consider the cyclic transformation of a piston moving up and down, the process will be the more irreversible the quicker we move the piston. Analogously in eternal inflation, we can think of de Sitter space as an equilibrium state, and inflation as the irreversible process associated to moving the scalar field along the potential. 
The slower the inflaton moves, the more inflation comes closer to becoming reversible and also closer to de Sitter space. Let us indeed consider the  equality $\Delta S_{ds}= 12\Omega N_c$ with $N_c$ the classical number of $e$-foldings, that is described later in the text. This equality is usually taken to represent the second law of thermodynamics $\Delta S=\left(\delta Q/T\right)_{rev}$, with $\left(\delta Q/T\right)_{rev}=12\Omega N_c$~\cite{Frolov:2002va,Jacobson:1995ab}. This interpretation is a bit puzzling to us, because it may look like that all classical solutions of General Relativity are reversible. On the other hand, the presence of an horizon suggests the presence of some irreversibility. In particular, slow roll inflation is irreversible at the classical level, and the flatter we make the potential, the closer we should get to reversibility. An interpretation of our formulas that would be more coherent with this intuition would be to use the bound $\Omega\geq 1$ for the non-eternally-inflating potentials, to say that whenever we do not have eternal inflation, $\Delta S\geq 12 N_c$, with the inequality being saturated at the phase transition to eternal inflation $\Omega=1$. It would be very interesting to establish such a connection between slow-roll inflation and thermodynamics on solid grounds, for 
it may give deeper understanding about de Sitter space and the onset of the eternal regime.


\section{Mini-review of known results}
It has been shown in \cite{Dubovsky:2008rf} that many information about the phase transition from non-eternal to
eternal slow-roll inflation are encoded in a rather simple formula. We briefly review here the results of that paper,
which we refer to for details.

In fact the probability distribution $\rho(V,\tau)$ for the volume
of the reheating surface $V$, with $\phi=H\tau/(2\sqrt6 \pi)$ the starting value for the inflaton field,
is given by
\begin{equation} \label{eq:rho1}
\rho(V,\tau)=\frac{1}{2\pi i}\int^{0^++i\infty}_{0^+-i\infty}dz\,f(\tau,z)e^{zV}\,,
\end{equation}
where at leading order in the slow-roll approximation, $f$ is the solution of 
the following differential equation (see also \cite{Winitzki:2008jp})
\begin{equation} \label{eq:diffeq1}
\partial_\tau^2 f(\tau;z)-2\sqrt\Omega \partial_\tau f(\tau;z)+f(\tau;z)\log[f(\tau;z)]=0\,,
\end{equation}
with
\begin{equation}
\Omega=\frac{2\pi^2}{3}\frac{\dot\phi^2}{H^4}\,,
\end{equation}
and  boundary conditions
\begin{align}
\label{eq:single_reheatinb_bc}f(0;z)=&e^{-z}\,,\\
\partial_\tau f(\tau;z)|_{\tau=\tau_b}=&0\,.
\end{align}
The first boundary condition corresponds to the end of inflation at $\tau=\phi=0$, while the second condition at $\tau=\tau_b$
is a barrier condition, to make finite the allowed field space for the inflaton\footnote{The barrier point $\tau_b$ can be thought of as
the region where the energy density of the inflaton potential becomes Planckian, or where
the potential becomes very steep, or the fixed point of a symmetric potential, such as the peak
in the top-of-the-hill inflationary model. The limit of arbitrary far barrier, $\tau_b\to\infty$, can be subtle and special care must be put in doing
such a limit, see~\cite{Creminelli:2008es,Dubovsky:2008rf} for details.}. 

The meaning of the differential equation is more manifest when rewritten in terms of $\phi$ as follows:
\begin{equation} \label{eq:diffeq2}
\frac12 \frac{\Delta \phi^2}{\Delta t} \frac{\partial^2}{\partial \phi^2} f
-\dot \phi \frac{\partial}{\partial \phi} f
+3 H f \log f=0\,.
\end{equation}
This is a modified Fokker-Planck equation: 
the first term is the normal dispersion term due to the quantum fluctuations 
of the inflaton field in de~Sitter space ($\Delta\phi^2/\Delta t=H^3/(4\pi^2)$); 
the second term is the drift induced by the tilt of the scalar potential;
the last term encodes the volume growth from the de~Sitter expansion. 

When rewritten in terms of $\tau$, the differential equation (\ref{eq:diffeq1}) only depends on the single
dimensionless parameter $\Omega$---a combination of the rate of quantum fluctuations ($\Delta\phi^2/\Delta t$),
the classical rolling ($\dot\phi$) and the Hubble expansion ($3H$)---which
controls the different phases of slow-roll inflation.

Despite an analytic expression for the solution to eqs.~(\ref{eq:rho1}) and (\ref{eq:diffeq1}) for $\rho(V;\tau)$ is not available, the
behaviors for the different regions of the parameters $V$ and $\Omega$ can be derived. Moreover all the moments, given by 
\be\label{eq:moments_single}
\langle V^n \rangle=(-1)^n \partial_z^n f(\tau;z)|_{z=0}\ ,
\ee 
can be computed
analytically, as eq.~(\ref{eq:diffeq1}) becomes linear for these quantities.

For example the expression for the average volume in terms of the classical number of $e$-foldings
$N_c=H\phi/|\dot\phi|$ reads
\begin{equation} \label{Vav1}
\langle V\rangle=e^{(\sqrt{\Omega}-\sqrt{\Omega-1})\tau}=e^{3N_c\frac{2}{1+\sqrt{1-1/\Omega}}}\,  , \qquad \Omega\geq 1.
\end{equation}

In the classical limit, $\Omega\to\infty$, quantum fluctuation become irrelevant, and $\rho(V;\tau)$
approaches a delta-function picked around the classical value $V_c=e^{3N_c}$.
When $\Omega\to1$ quantum fluctuations become of order one and the average volume gets large corrections,
increasing to $\langle V\rangle=e^{6N_c}$. At the critical value $\Omega=1$, the phase transition to eternal inflation
occurs and the average volume starts diverging.


\section{Universality of the Bound in $D$-dimensions}
The first test of the eternal inflation bound we present here is its universality on the number of space-time dimensions.
We remind that the bound follows from two ingredients: a classical one, which determines the largest non-eternally inflating 
classical trajectory, and a quantum one, which compute the deformation due to quantum fluctuations. As described in the previous section
the second computation boils down to solve a differential equation and performing a Laplace transformation.
The transition to eternal inflation happens when the only parameter of the differential equation $\Omega$ crosses 1.
The expression for the differential equation given in eq.~(\ref{eq:diffeq2}) is also valid in $D$ space-time dimensions
except for the coefficient of the last term, the Hubble expansion coefficient, which now becomes $(D-1)H$.
The expression for $\Delta\phi^2/\Delta t$ in $D$-dimensions is also different, it can be extracted from
the coefficient of the linear term in the 2-point function of a scalar field in dS$_D$,
\begin{equation} \label{eq:2pointsdS}
\langle \phi^2 \rangle=\frac{H^{D-1}}{\pi \Theta_{(D-1)}}t+\dots
\end{equation}
where 
$$\Theta_{(d)}=\frac{2\pi^{d/2}}{\Gamma(\frac d 2)}$$
is the $d$-dimensional solid angle.
The differential equation can be easily brought back to the form~(\ref{eq:diffeq1}), by defining
\begin{align} \label{eq:DdepOm}
\Omega=&\frac{\dot\phi^2}{2(D-1)H}\frac{\Delta t}{\Delta\phi^2}=\frac{\pi\Theta_{(D-1)}}{2 (D-1)}\frac{\dot\phi^2}{H^D}\,, \\
\tau=&\phi \sqrt{2(D-1)H} \sqrt{\frac{\Delta t}{\Delta \phi^2}}=2(D-1) \sqrt\Omega N_c\,.
\end{align}

Since the differential equation governing quantum fluctuations has the same form in $D$ dimensions,
the same will be true for the solution when expressed in term of $\tau$. In particular the expression for the average volume will
be
\begin{equation} \label{eq:VmedD}
\langle V \rangle=e^{(D-1)N_c \frac{2}{1+\sqrt{1-1/\Omega}}}\,,
\end{equation}
with $\Omega$ as defined in eq.~(\ref{eq:DdepOm}). Again for $\Omega\to\infty$ the classical result
is recovered $\langle V\rangle=e^{(D-1)N_c}$, while the value at the eternal inflation transition (at $\Omega=1$
as eq.~(\ref{eq:diffeq1}) is formally unchanged), 
is $\langle V\rangle=e^{2(D-1)N_c}$, again quantum fluctuations increase the effective number of $e$-foldings
up to a factor 2 with respect to the classical value when the phase transition is approached.

In order to test the bound we need now to work out the relation between the classical number of $e$-folding
and the de~Sitter entropy in $D$-dimensions. The difference of de~Sitter entropy between the start and the end
of inflation can be written as
\begin{align}
\Delta S=\int_{S_{start}}^{S_{end}} dS=\int_{A_{start}}^{A_{end}} \frac{1}{4G}dA= -\int_0^{N_c} \frac{(D-2)}{4G}\frac{\Theta_{(D-1)}\dot H}{H^{D}} dN'_c\,,
\end{align}
where we used
$$A=\frac{\Theta_{(D-1)}}{H^{D-2}}$$ for the de~Sitter horizon area.
From Friedmann equations in $D$ dimensions we also have
\begin{equation}
\dot H=-\frac{8\pi G}{D-2}\dot\phi^2\,,
\end{equation}
which allows us to write
\begin{equation}
\Delta S=\int_0^{N_c} 2\pi \Theta_{(D-1)}\frac{\dot\phi^2}{H^D}dN'_c.
\end{equation}
We can now use eq.~(\ref{eq:DdepOm}) to rewrite $\dot\phi$ in terms of $\Omega$ and we finally get
\begin{align}
\Delta S&=\int_0^{N_c} 4(D-1)\Omega dN_c=4(D-1)\Omega N'_c\,,\\
S_{end}&\geq 4(D-1)\Omega N_c\,,
\end{align}
where we used the fact that at leading order in the slow-roll parameter $\Omega=$const.
The result above gives a bound on the number of classical
$e$-foldings allowed in non-eternal inflation when $\Omega>1$. The maximum number is achieved
at the phase transition $\Omega=1$, where $N_c\leq S_{end}/(4(D-1))$.
Quite non-trivially the complicated dependence on the number of dimensions $D$ simplifies considerably
in the final expression and only a $D-1$ factor remains. 

We can now combine this information with the full quantum computation of eq.~(\ref{eq:VmedD}),
we thus have
\begin{equation}\label{eq:Vavbound}
\langle V \rangle=e^{(D-1)N_c \frac{2}{1+\sqrt{1-1/\Omega}}}
\leq e^{\frac{S_{end}}{2} \frac{1}{\Omega(1+\sqrt{1-1/\Omega})}}\leq e^{\frac{S_{end}}{2}}\,, \qquad {\rm for}\  \Omega\geq1.
\end{equation}

Remarkably, once written in terms of the de~Sitter entropy, the bound on the volume is universal, 
independent of the number of dimensions! As in $D=4$ the probability to produce a finite volume
violating the bound above is super-exponentially small (see \cite{Dubovsky:2008rf} for details), i.e. zero 
within the effective field theory regime.

\section{Universality of the Bound with Higher-Derivative Corrections}
\label{sec:hd}
The second test of the bound that we provide is with respect to higher-derivative corrections in the Einstein-Hilbert plus inflaton action. Higher derivative terms have multiple effects: they modify the Einstein equations, the expression for the energy-momentum tensor, the inflaton equations of motion and the formula for the entropy as well. 
Since we are interested here in slow-roll inflation we will assume that independently of the modification induced by higher-derivative terms there exists a solution where the metric is approximately de~Sitter up to slow-roll corrections
and the inflaton field rolls slowly, i.e. $\ddot \phi\ll H\dot \phi$.

This assumption may appear somewhat too restrictive. There are inflationary models (such as DBI~\cite{Alishahiha:2004eh} and ghost \cite{ArkaniHamed:2003uz}
inflation) where the scalar field does not roll slowly even though the geometry is approximately de~Sitter. However, as shown in \cite{ArkaniHamed:2007ky}, these models are very far from 
saturating the volume bound as soon they obey the null energy condition and do not exhibit superluminal excitations. On the other hand, ghost inflation is capable of violating the bound even at the classical level at the expense of violating the null energy condition.
Given that the null energy condition and the absence of superluminal excitations are crucial for the success of black hole thermodynamics \cite{Dubovsky:2006vk}
we consider this link as yet another indication that the volume bound has a thermodynamical origin. We therefore restrict ourselves to models that are perturbatively close to slow roll inflation.

The most general action for the graviton plus inflaton system can be written as
\begin{align}
{\cal L} &={\cal L} (g_{\mu\nu},\, R_{\mu\nu\rho\sigma}, \nabla_\alpha R_{\mu\nu\rho\sigma},...,\phi,\nabla_\alpha \phi,...) \nonumber \\
&={\cal L}_{G}(g_{\mu\nu},\, R_{\mu\nu\rho\sigma}, \nabla_\alpha R_{\mu\nu\rho\sigma}, ...,\phi)
	+{\cal L}_{kin}(g_{\mu\nu},R_{\mu\nu\rho\sigma},\nabla_\alpha R_{\mu\nu\rho\sigma},...,\phi,\nabla_\alpha \phi,...)\,,
\end{align}
where we have isolated the part of the Lagrangian that does not depend on $\nabla_\alpha \phi$ (${\cal L}_G=\frac{1}{16\pi G}R-V(\phi)+\dots$)
 from the one that contains derivatives of the inflaton (${\cal L}_{kin}=-\frac12 (\partial \phi)^2+\dots$).

We immediately realize that at leading order in slow-roll parameters we can neglect the contribution of the terms in ${\cal L}_G$ involving covariant derivatives. This is justified in the leading slow-roll approximation after realizing that all the terms of this form are at least of second order in the slow-roll parameters. Indeed all such terms must involve at least two derivatives in order to contract the indexes. Upon integration by parts, it is therefore possible to have at least two of these derivatives acting on different Riemann tensors. The Riemann tensor in de Sitter space is proportional to the metric and so covariantly constant, hence its covariant derivatives in an inflationary spacetime are proportional to the slow roll parameters. This means that terms in ${\cal L}_G$ involving covariant derivatives start at second order in the slow roll parameters and can therefore be neglected in our approximation. 

For similar reasons at leading order in the slow-roll expansion we will only consider up to 2-derivative terms 
acting on the inflaton, i.e.
\begin{equation} \label{eq:Lm}
{\cal L}_{kin}(g_{\mu\nu},R_{\mu\nu\rho\sigma},...,\phi,\nabla_\alpha \phi,...)=-\frac12(\partial_\alpha \phi \partial_\beta \phi) \Pi^{\alpha\beta}(g_{\mu\nu}, R_{\mu\nu\rho\sigma},...,\phi)\,.
\end{equation}
In the same approximation, since $(\partial \phi)^2$ is already subleading in the slow-roll expansion, its coefficient
can be taken at 0-th order in slow-roll, i.e. computed with de~Sitter metric, this gives
\begin{equation}
\Pi^{\alpha\beta}(g_{\mu\nu},R_{\mu\nu\rho\sigma},...)=g^{\alpha\beta} \Pi(g_{\mu\nu},\,R_{\mu\nu\rho\sigma},...)
\end{equation}
where $\Pi=\Pi^{\mu\nu}g_{\mu\nu}/4$.

We can now proceed to write the Einstein equations.
It is possible to show that ignoring higher derivatives the gravitation Lagrangian can be written simply in terms of $R^{\mu\nu}{}_{\rho\sigma}$ and 
$R^{\mu}{}_{\nu}{}^{\rho}{}_{\sigma}$ without any explicit dependence on the metric:
\begin{equation}
{\cal L}_{G}(g_{\mu\nu},\, R_{\mu\nu\rho\sigma}, ...,\phi)=\widetilde {\cal L}_{G}(R^{\mu\nu}{}_{\rho\sigma}, \, R^{\mu}{}_{\nu}{}^{\rho}{}_{\sigma}, ...,\phi)\,,
\end{equation}
and analogously for $\Pi$:
\begin{equation}
{\Pi}(g_{\mu\nu},\, R_{\mu\nu\rho\sigma}, ...,\phi)=\widetilde {\Pi}(R^{\mu\nu}{}_{\rho\sigma}, \, R^{\mu}{}_{\nu}{}^{\rho}{}_{\sigma}, ...,\phi)\,.
\end{equation}
This is proven in appendix~\ref{app:one}.
In this way the equations of motion for the metric read
\begin{equation} \label{eq:eegen}
\frac{\delta \widetilde{\cal L}}{\delta R^{\alpha\beta}{}_{\rho\sigma}}\frac{\delta R^{\alpha\beta}{}_{\rho\sigma}}{\delta g^{\mu\nu}}+
\frac{\delta \widetilde{\cal L}}{\delta R^{\alpha}{}_{\beta}{}^{\rho}{}_{\sigma}}\frac{\delta R^{\alpha}{}_{\beta}{}^{\rho}{}_{\sigma}}{\delta g^{\mu\nu}}
-\frac12 g_{\mu\nu} \widetilde{\cal L}
=\frac12 \widetilde \Pi \partial_{\mu}\phi \partial_{\nu}\phi\,.
\end{equation}
Using the relation
\begin{equation} \label{eq:waldid}
\frac{\delta {\cal L}}{\delta R_{\mu\nu\rho\sigma}} \delta R_{\mu\nu\rho\sigma}=
\frac{\delta {\cal L}}{\delta R_{\mu\nu\rho\sigma}}\Bigl(2 \nabla_\mu\nabla_\sigma \delta g_{\nu\rho}+R^{\tau}{}_{\nu\rho\sigma}\delta g_{\mu\tau} \Bigr)\,,
\end{equation}
it is possible to rewrite the equations of motion as (see appendix~\ref{app:one} for details):
\begin{equation} \label{eq:eegen2}
-2 \nabla_{\alpha}\nabla_{\beta}\frac{\delta {\cal L}}{\delta R_{\alpha(\mu\nu)\beta}}
+R^{(\mu}{}_{\alpha\beta\gamma}\frac{\delta {\cal L}}{\delta R_{\nu)\alpha\beta\gamma}}
-\frac12 g^{\mu\nu} {\cal L}
=\frac12 \Pi \partial^{\mu}\phi \partial^{\nu}\phi
\end{equation}
where indices in between brackets are symmetrized.

From here we can see that higher derivative terms change non trivially several relations used for the proof
of the bound. The Friedmann equation $\dot H=-4\pi G \dot \phi^2$ used to relate the change in entropy
with the $\Omega$ parameter gets modified, in particular both the l.h.s. and the r.h.s. receive corrections.
The modification of the inflaton kinetic term changes the 2-point function of the inflaton in the de~Sitter
phase, which corresponds to a change in the expression of $\Omega$. 
Finally also the expression of the entropy changes, since for a generic gravity Lagrangian
the Wald formula~\cite{Wald:1993nt,Iyer:1994ys} must be employed:
\begin{equation} \label{eq:waldS}
S=-\frac{4\pi}{\kappa}\int_{{\cal H}}d\Sigma_{\mu\nu}Q^{\mu\nu}
\end{equation}
where $\kappa$ is the surface gravity on the horizon ${\cal H}$,
$d\Sigma_{\mu\nu}=\frac12 \epsilon_{\mu\nu}dA$ is the area element ($\epsilon_{\mu\nu}$ is
a tensor binormal to ${\cal H}$ normalized such that $\epsilon_{\mu\nu}\epsilon^{\mu\nu}=-2$) and
\begin{equation}
Q^{\mu\nu}=\frac{\delta {\cal L}}{\delta R_{\mu\nu\rho\sigma}}\nabla_\rho \xi_{\sigma}
-2\nabla_{\rho}\frac{\delta {\cal L}}{\delta R_{\mu\nu\rho\sigma}} \xi_{\sigma}\,,
\end{equation}
where $\xi_{\mu}$ is the Killing vector, which on the horizon satisfies the relation $\nabla_{\mu}\xi_\nu=\kappa \epsilon_{\mu\nu}$.

We can now proceed to calculate the variation of the entropy
in analogy to what has been done in the black hole case \cite{Parikh:2009qs} or
in slow-roll for Einstein gravity \cite{Frolov:2002va}. 
\begin{figure}[t]
\begin{center}
\includegraphics[height=0.36\textwidth]{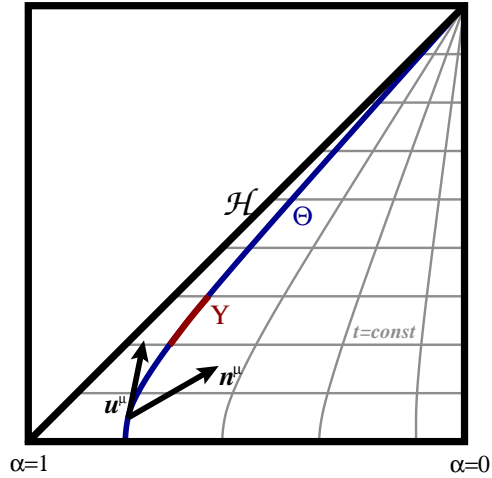} 
\caption{\label{fig:dSfrw} \small \it The stretched horizon (in blue) in de~Sitter space, with the unit vectors
$u^\mu\propto \xi^\mu$ and $n^\mu$ and the region $\Upsilon$ (in red) spanned by the area section $\Sigma(t)$,
as defined in the text. 
}
\end{center}
\end{figure}
We start defining (see fig.~\ref{fig:dSfrw}) a stretched horizon $\Theta$ inside the true de~Sitter horizon $\cal H$
(in FRW coordinates it would correspond to the region of points with $r=\alpha H^{-1} e^{-Ht}$, with $\alpha<1$).
Within $\Theta$ the Killing vector $\xi^\mu=(1,-H r,0,0)$ has constant norm $\xi_\mu \xi^\mu=-(1-\alpha^2)$.
The norm is zero on the horizon, since $\xi^\mu$ becomes null, and normalized to -1 on the origin $r=\alpha=0$,
where the FRW observer sits.
$\Theta$ is also characterized by two unit vectors $u_\mu=\xi_\mu/(1-\alpha^2)^{1/2}$ parallel to the Killing vector and $n^\mu=(\alpha,-H r/\alpha)/(1-\alpha^2)^{1/2}$, which is orthogonal to $\Theta$ and pointing away from the horizon. 
In the limit where $\Theta$ approaches $\cal H$, $\alpha\to1$ and both $u^\mu$
and $n^\mu$ become proportional to $\xi^\mu$, which becomes null.
On the stretched horizon,   at each moment in time $t$, we can use  these two vectors to define 
the area element $d\Sigma_{\mu\nu}=\frac12(n_\mu u_\nu-n_\nu u_\mu)dA\equiv\frac12 \epsilon_{\mu\nu} dA$
for the 2-sphere $\Sigma(t)$, which is the constant-$t$ section of $\Theta$.
Moving along $\xi_\mu$, $\Sigma$ spans a three-volume $\Upsilon$ where $\kappa$ is constant
and the variation of the entropy can be computed using Stokes theorem:
\begin{equation}
-\delta \frac{4\pi}{\kappa}\int_{\Sigma(t)} d\Sigma_{\mu\nu} Q^{\mu\nu}=
-\frac{4\pi}{\kappa}\oint_{\partial\Upsilon} d\Sigma_{\mu\nu} Q^{\mu\nu}=
\frac{4\pi}{\kappa}\int_{\Upsilon} d\tau dA n_\mu \nabla_{\nu} Q^{\mu\nu}.
\end{equation}
where $d\tau$ is the unit (proper) time interval in $\Theta$.

After few algebraic manipulations (see appendix~\ref{app:one}) we arrive at the following formula:
\begin{equation} \label{eq:dsdtinter}
\frac{dS}{dt}=\frac{2\pi}{\kappa}\int_{{\cal H}}dA \, \xi_{\mu} \xi_\nu \left( 
\frac{\delta {\cal L}}{\delta R_{\mu\alpha\beta\gamma}} R^{\nu}{}_{\alpha\beta\gamma}
-2\nabla_{\alpha}\nabla_{\beta}\frac{\delta {\cal L}}{\delta R_{\mu\alpha\beta\nu}}\right)\,.
\end{equation}
After  using the equations of motion (\ref{eq:eegen2}) we have
\begin{equation} \label{eq:dSdt1gen}
\frac{dS}{Hdt}=\frac{2\pi}{\kappa}\frac{A({\cal H})\Pi}{H}\, \partial_{\mu}\phi \partial_{\nu} \phi \, \xi^\mu \xi^{\nu}=
12\, \Pi\,  \frac{2\pi^2}{3} \frac{\dot\phi^2}{H^4}\,.
\end{equation}

The first non trivial result is that in eq.~(\ref{eq:dSdt1gen}) the non-trivial dependence on the Riemann tensor
of the l.h.s. of the Einstein equation (\ref{eq:eegen}) is completely gone after using the Wald formula for the entropy.
We will now show that the residual dependence in the r.h.s. will also disappeared after
taking into account the modification of the inflaton action.
Indeed the kinetic term for the inflaton in eq.~(\ref{eq:Lm}) is not canonical anymore.
This implies that the corresponding 2-points function during the de~Sitter phase now reads
\begin{equation} 
\langle \phi^2 \rangle=\frac{H^3}{4\pi^2 \Pi}t+\dots
\end{equation}
The corresponding definition (\ref{eq:DdepOm}) for $\Omega$  will thus be
\begin{equation} \label{eq:Omgen2}
\Omega=\frac{\dot\phi^2}{6H}\frac{\Delta t}{\Delta\phi^2}=
 \Pi \, \frac{2\pi^2}{3}\frac{\dot\phi^2}{H^4}\,,
\end{equation}
which finally allows us to write eq.~(\ref{eq:dSdt1gen}) again in the form
\begin{equation}
\frac{dS}{dN}=12\,\Omega\,.
\end{equation}
The differential equation governing quantum fluctuations of the inflaton will also have the same form as 
in the Einstein gravity case once $\Omega$ is properly defined according to eq.~(\ref{eq:Omgen2}).

We thus conclude that, in the approximation we are working, i.e. at leading order in the slow-roll parameter,
higher derivative corrections in the Lagrangian do not modify the calculation of the probability distribution
of the volume  of the reheating surface after inflation. In particular the bound on the volume of the universe after
slow-roll inflation is universal also with respect to higher derivative corrections---the coefficient ``$1/2$''
in the exponent of eq.~(\ref{eq:Vavbound}) does not receive corrections from higher-derivative terms!


\section{Universality of the Bound in Multifield Inflation}

After having seen that the bound on the finite volume holds in any number of dimensions and after including higher derivative terms in the Eistein-Hilbert action, we now pass to the study of the case where we have more than one light field during inflation: multifield inflation.

\begin{figure}[t]
\begin{center}
\includegraphics[height=0.36\textwidth]{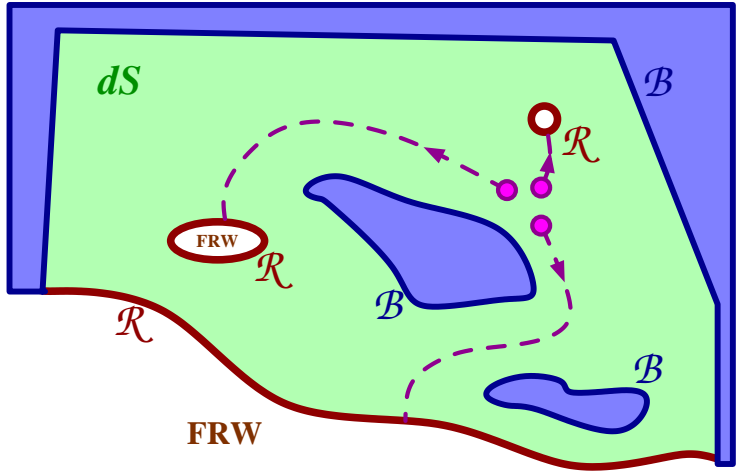} 
\caption{\label{fig:potgen} \small \it Depending on the starting point, inflaton trajectories in field space (in magenta) drift towards 
different points of the reheating surface (in red), which in general correspond to different vacua. 
The field space is often confined by boundary regions (in blue) where, for instance, the energy density of the scalar potential become Planckian.
}
\end{center}
\end{figure}

It is straightforward to generalize the procedure of ref.~\cite{Dubovsky:2008rf} that led to eq.~(\ref{eq:diffeq1}) to the case of multifield inflation. 
We now have
\begin{equation}\label{eq_multi_single_reheating}
\bm{\nabla}^2 f({\bm{\tau}},z)-2{\bm{\sqrt{\Omega}}}\cdot{\bm{\nabla}} f(\bm{\tau},z)+f(\bm{\tau},z)\log[f(\bm{\tau},z)]=0\,,
\end{equation}
where
\be
\bm{\nabla}=\partial_{\bm{\tau}}\ , \qquad
\bm{\tau}=2\pi\sqrt6 \frac{\bm{\phi}}{H}\ , \qquad
\bm{\sqrt{\Omega}}=\sqrt\frac{2\pi^2}{3} \frac{\bm{\dot\phi}}{H^2}\ ,
\ee
the boldface font is used for vectors in field space, and we assumed  for the moment that the reheating surface is just a single point at $\bm{\tau}=\bm{\tau}_{r}$. We assume there is also a barrier ${\cal B}$ that bounds the moduli space. In this case the generating function must satisfy the boundary conditions 
\bea\label{eq:single_point_reheatinb_bc}
&f(\bm{\tau}_r,z)=e^{-z}\,, \\ \nonumber
&\left.\bm{\hat n}\; \cdot \bm{\nabla} f(\bm{\tau}, z)\right |_{\bm{\tau}\in {\cal B}}=0\ .
\eea
where $\hat n$ is the normal versor to the barrier hypersurface.
This is a simple generalization of the analogous boundary condition for single field inflation eq.~(\ref{eq:single_reheatinb_bc}).
The probability distribution for the volume is given by
\be\label{eq:volume_single_rehating}
\rho(V,\bm{\tau}) =\int_{0^+-i\infty}^{0^++i\infty} d z\; e^{z V}\, f(\bm{\tau}, z)\ .
\ee
More generally the reheating region may be a surface in field space ${\cal R}$, as depicted in fig.~\ref{fig:potgen}.
In this case different points on $\cal R$ might correspond to local universes with different physical properties. Hubble patches terminating inflation
on different points of the reheating region may have different Hubble constants, entropy, and even correspond to
different vacua (e.g.~if $\cal R$ is disconnected or it represents a moduli space).
In this situation it is not clear the meaning of the bound on the volume relative to the entropy: the entropy may vary substantially from point to point in $\cal R$, and it is not obvious which of the entropies associated to the various vacua we should take. We may consider the largest one as an upper bound, but, as we are now going to see, we can have a more stringent definition. 

In fact, it is possible to keep track of the \emph{type of volume} produced at the end of inflation by tracking the point where the fields exit inflation.
If we assume that the reheating surface $\cal R$ is made of a set of $n$ disjoint points ($\bm{\tau}_r^i$) in field space, the boundary condition at $\cal R$ for $f$ would read
\begin{equation}
f(\bm{\tau}_r^i,\vec z)=e^{-z_i}\, ,
\end{equation}
generalizing the boundary condition of eq.~(\ref{eq:single_point_reheatinb_bc}) by using a different variable $z_i$ for each reheating point $\bm{\tau}_r^i$.
The conjugate variable to each $z_i$ would be $V_i$, the volume of type $\bm{\tau}_r^i$.
The probability distribution of creating the volumes $\vec V=\{V_1,\dots,V_n\}$ of type $\vec{\bm{\tau}}_r=\{\bm{\tau}_r^1,\dots,\bm{\tau}_r^n\}$ respectively, is then the $n$-dimensional Laplace anti-transform
\begin{equation}
\rho(\vec V,\vec{\bm{\tau}})=\int_{{\cal C}} d \vec z\; e^{\vec V \cdot \vec z}\, f(\vec{\bm{\tau}},\vec z)\ ,
\end{equation}
where ${\cal{C}}=\{\vec z: z\in \mathbb{C}\ \wedge \  {\rm Re}(z_i)=0^+ \}$  simply generalizes the contour of eq.~(\ref{eq:rho1}).

In the case of a continuos reheating surface in field space the vector $\vec{\bm{\tau}}_r$ becomes a continuous variable $\bm{\tau}_r\in{\cal R}$,
$\vec{z}$ and $\vec{V}$ become functions of $\bm{\tau}_r$ respectively $z(\bm{\tau}_r)$ and $V(\bm{\tau}_r)$, and the formula for
the probability distribution of the volume becomes a functional integral
\begin{equation}
\label{eq:multi_laplace}
\rho(V(\bm{\tau}_r),\bm{\tau})=\int_{\cal C} {\cal D} z(\bm{\tau}_r)\; e^{\int_{\cal R} d\bm{\tau}'_r  V(\bm{\tau}'_r) z(\bm{\tau}'_r)}\; 
	f(\bm{\tau}, z(\bm{\tau}_r))\,,
\end{equation}
with $\bm{\tau}_r$ spanning $\cal R$ and normalized so that $\int_{\cal R} d\bm{\tau}_r=1$, and $f(\bm{\tau}, z(\bm{\tau}_r))$ now satisfying
\begin{align}
\label{eq:multi-system}
 \bm{\nabla}^2 f(\bm{\tau},z(\bm{\tau}_r))
-2\bm{\sqrt{\Omega}}\cdot\bm{\nabla} f(\bm{\tau},z(\bm{\tau}_r))&+f(\bm{\tau},z(\bm{\tau}_r))\log[f(\bm{\tau},z(\bm{\tau}_r))]=0\,, \\ \nonumber
f(\bm{\tau}_r,z(\bm{\tau}_r))&=e^{-z(\bm{\tau}_r)}\,, \\ \nonumber
\left. \bm{\hat n}\; \cdot\bm{\nabla}  f(\bm{\tau}, z(\bm{\tau}_r))\right |_{\bm{\tau}\in {\cal B}}&=0\ .
\end{align}

It is possible at this point to show that eq.~(\ref{eq_multi_single_reheating}) does not apply only to the case where we have a single reheating point. It is also the generating function for the probability distribution of the total volume $V=\int_{\cal R}d\bm{\tau}_r\;V(\bm{\tau}_r)$ produced at the end of inflation. Indeed the probability distribution for the total volume $V$ is given by~\footnote{The actual derivation reads
\begin{align} \nonumber
\rho(V,\bm{\tau})&=\int {\cal D} V(\bm{\tau}_r)\; \rho(V(\bm{\tau}_r),\bm{\tau})\;\delta\left(\int_{\cal R} d\bm{\tau}'_r V(\bm{\tau}'_r)-V\right) =\int_{-\infty}^{\infty} d \lambda \int {\cal D} V(\bm{\tau}_r)\; 
		\rho(V(\bm{\tau}_r),\bm{\tau})\;e^{i \lambda (\int_{\cal R} d\bm{\tau}'_r V(\bm{\tau}'_r)-V)}\\ \nonumber 
	&=\int_{-\infty}^{\infty} d \lambda \int {\cal D} V(\bm{\tau}_r) 
		\int {\cal D} z(\bm{\tau}_r) \;
		e^{\int_{\cal R} d\bm{\tau}'_r V(\bm{\tau}'_r)  z(\bm{\tau}'_r)+i \lambda(\int_{\cal R} d\bm{\tau}'_r V(\bm{\tau}'_r)-V)}\; f(\bm{\tau}, z(\bm{\tau}_r)) \\\nonumber 
	&=\int_{-\infty}^{\infty} d \lambda 
		\int {\cal D} z(\bm{\tau}_r) \,
		e^{-i \lambda V} f(\bm{\tau}, z(\bm{\tau}_r))\; \delta(\lambda- i z(\bm{\tau}_r))=-i\int_{-\infty}^{\infty} d \lambda\, e^{-i \lambda V} f(\bm{\tau}, -i\lambda)=\int_{0^+-i\infty}^{0^++i\infty} d z\, e^{z V} f(\bm{\tau}, z)\ .
\end{align}}:

\begin{align}
\rho(V,\bm{\tau})&=\int {\cal D} V(\bm{\tau}_r)\; \rho(V(\bm{\tau}_r),\bm{\tau})\;\delta\left(\int_{\cal R} d\bm{\tau}'_r\; V(\bm{\tau}'_r)-V\right) =\int_{0^+-i\infty}^{0^++i\infty} d z\; e^{z V}\, f(\bm{\tau}, z)\ .
\end{align}
This is exactly eq.~(\ref{eq:volume_single_rehating}). Therefore we conclude that solving equation (\ref{eq_multi_single_reheating}) with boundary conditions (\ref{eq:single_point_reheatinb_bc}) corresponds to compute the generating function for the probability distribution of the total reheated volume, independently of the particular kind of volumes this is made of.

At this point the equation for the moments of the distribution (\ref{eq:moments_single}) can be simply generalized to the multifield case as follows
\be\label{eq:momenta}
\langle V(\bm{\tau}_r') V(\bm{\tau}_r'')\ldots  V(\bm{\tau}_r^{(n)})\rangle=(-1)^n\left.\frac{\delta^n f\left(\bm{\tau},z(\bm{\tau}_r)\right)}{\delta z(\bm{\tau}_r')\delta z(\bm{\tau}_r'')\ldots\delta z(\bm{\tau}_r^{(n)})}\right|_{z(\bm{\tau}_r)=0}\ .
\ee

The differential equations for the moments of the distributions are linear and in particular the one for the average volume reads
\begin{align}
 \bm{\nabla}^2 \langle V(\bf{\tau}'_r)\rangle &
-2\bm{\sqrt{\Omega}}\cdot\bm{\nabla} \langle V(\bf{\tau}'_r)\rangle +\langle V(\bf{\tau}'_r)\rangle =0 \\ \nonumber
&\langle V(\bf{\tau}'_r)\rangle|_{\bf{\tau}=\bf{\tau}_r} =\delta (\bf{\tau}'_r-\bf{\tau}_r) \,, \\ \nonumber
&\left. \bm{\hat n}\; \cdot\bm{\nabla}\langle V(\bf{\tau}'_r)\rangle \right |_{\bm{\tau}\in {\cal B}}=0\ .
\end{align}

\noindent
\vspace{0.2cm}
{\bf The bound:}\\
We are now ready to give a more stringent definition of the bound that we believe it will still be satisfied in general. In the case of multifield inflation, there are multiple kind of vacua. According to the meaning of the bound on the volume of the reheating surface in single field inflation, we expect that, in the non eternal inflation phase the total volume produced of any kind be bounded by the corresponding  de-Sitter entropy; in other words we expect a bound to exist for any particular kind of volume. We are therefore led to conjecture the following simple generalization of the bound on the volume of inflation:
\bea
&&P\left(\int_{\cal I}d\bm{\tau}_r\; V(\bm{\tau}_r)> {\rm Sup}_{\cal I}\left[e^{S(\bm{\tau}_r)/2}\right];V<+\infty\right)=\\ \nonumber
&&\quad \  =\int_{\cal I}d\bm{\tau}_r\int_{e^{S(\bm{\tau}_r)/2}\delta(\bm{\tau}_r'-\bm{\tau}_r)}^{+\infty}{\cal D}V(\bm{\tau}_r')\ \rho\left(V(\bm{\tau}_r',\bm{\tau})\right)\lesssim {\rm Sup}_{\cal I}\left[e^{-k\, e^{S(\bm{\tau}_r)/2}}\right]\ ,
\eea
for any subset ${\cal I}$ of ${\cal R}$ and with $k$ being a numerical factor of order one. The bound on the volume states that the probability to create a volume of kind $\bm{\tau}_r$ larger than $e^{S_{dS}/2}$ and total volume finite, vanishes up to non-perturbatively small quantum gravity effects. 

In particular, for the average of the volume this implies
\be\label{eq:average-bound}
\int_{\cal I}d\bm{\tau}_r\;  \langle V(\bm{\tau}_r) \rangle\lesssim {\rm Sup}_{\cal I}\left[e^{S(\bm{\tau}_r)/2}\right]\ ,
\ee
and similarly for higher moments.

Given a certain multifiled inflationary model, one can solve eq.~(\ref{eq:multi-system}) and obtain the probability distribution for any kind of volume after performing the (functional) inverse Laplace transform~(\ref{eq:multi_laplace}). In practice, such a task is extremely difficult from a technical point of view. What we will do next is presenting two simple examples where we are able to compute explicitly $\langle V(\bm{\tau}_r)\rangle$ and that offer a non-trivial check for the bound in~(\ref{eq:average-bound}). In fact checking for the average
is expected to be enough because as it has been shown in the single field case the probability distribution 
is always sharply peaked around the average value.

\subsection{First example: the waterfall}


\begin{figure}[t]
\begin{center}
\includegraphics[height=0.36\textwidth]{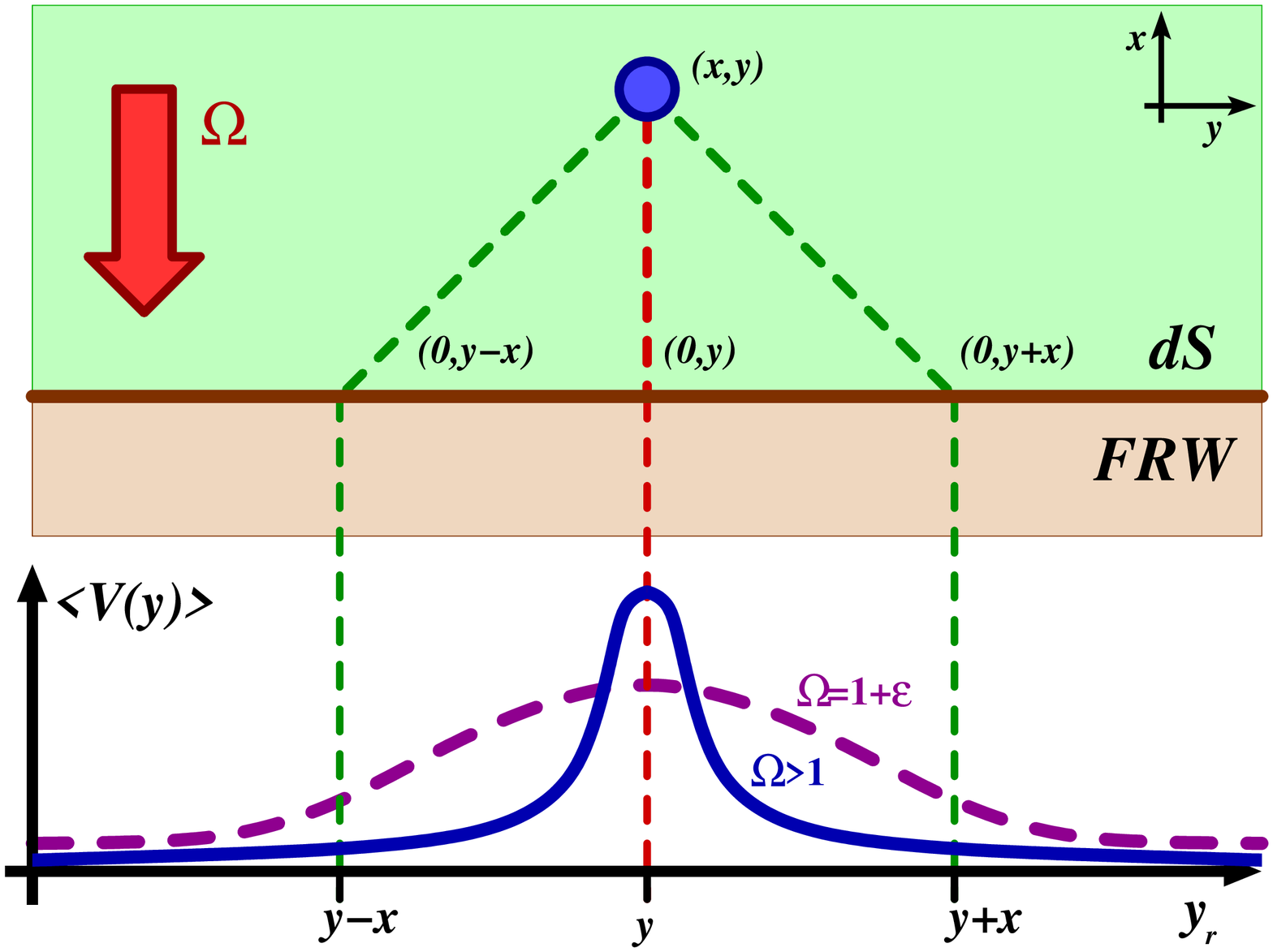} 
\includegraphics[height=0.36\textwidth]{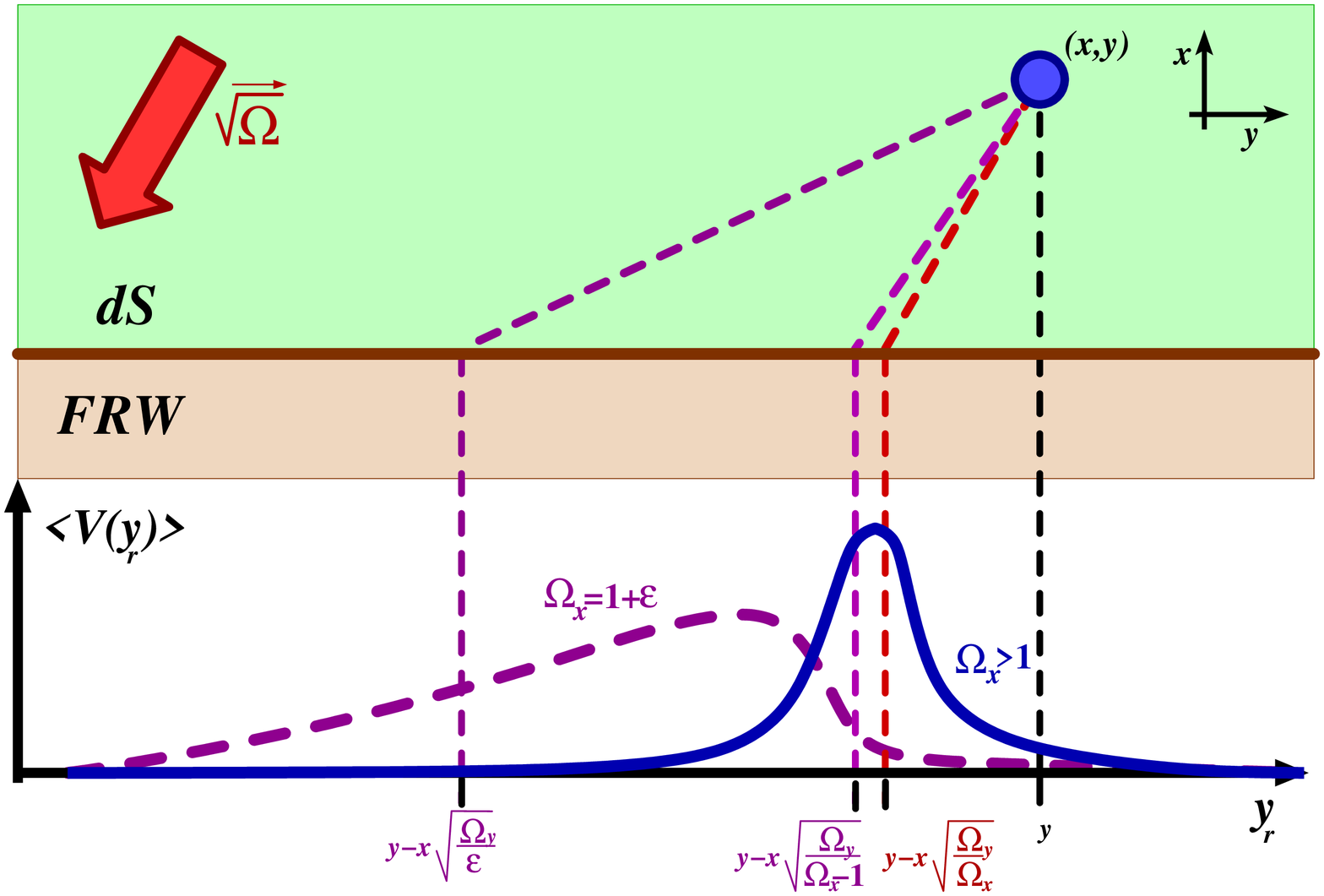} 
\caption{\label{fig:waterfall} \small \it Average volume distribution  $\langle V(y_r)\rangle$ as a function of $y_r$ for
the two inflationary examples: the waterfall \emph{(left)} and the tilted waterfall \emph{(right)}.
The red-dashed lines refer to the classical evolution. Near the classical limit, for large $\Omega$, 
the distribution (in blue) is peaked around the classical exit point. Near the phase transition to eternal inflation, 
at $\Omega_x \gtrsim 1$, the distribution (in magenta) broadens (left) and drifts towards smaller values of $y_r$ (right).
}
\end{center}
\end{figure}

The first and simplest example we consider is the two-field case $\bm{\tau}=(x,y)$ where the slope of the potential is uniform everywhere and orthogonal to the reheating surface at $\bm{\tau}_r=(0,y_r)$, i.e. ${\bm{\sqrt\Omega}}=(\sqrt{\Omega},0)$. Physically this corresponds to the inflationary model where the second field acts just as a classically irrelevant spectator (Fig.~\ref{fig:waterfall}).  In this case the two fields decouple because of the shift symmetry in $y$, and we expect the result to be quite close to the single field case. In particular, we expect to recover the same result as in single field inflation once we integrate over the position of the reheating point. Though quite similar to the single field case, this example is simple enough to allow us to calculate explicitly also the distribution of ``different volumes'' on the reheating surface. This will lead us to quite non-trivial results.

For example let us calculate the average volume of type $y_r$ given the starting point~$\tau=(x,y)$. Applying (\ref{eq:momenta}) we obtain
\begin{equation}
\langle V(y_r) \rangle=-\left.\frac{\delta f(x,y;z(y_r'))}{\delta z(y_r)}\right|_{z(y_r')=0}\equiv - \psi(x,\Delta y)\ .
\end{equation}
where we have used the shift symmetry in $y$ to assume that the $y$-dependence can be only in the form $\Delta y=y-y_r$. $\psi(x,\Delta y)$ satisfies
\begin{align}\label{eq:master-local}
\psi_{xx}&+\psi_{yy}-2\sqrt\Omega \psi_{x}+\psi=0\ ,\\ \nonumber
&\left.\psi(0,\Delta y)=-\delta(\Delta y)\right. \ , \\ \nonumber
&\left. \psi_x(x,\Delta y)\right|_{x=x_b}=0\ ,
\end{align}
where $x_b$ is the $y$-independent location of the barrier.

The solution reads
\begin{align}
\psi(x,\Delta y)&=\int_{-\infty}^{\infty} \frac{dk}{2\pi}e^{i\,k\, \Delta y}\frac{\omega_{k-} e^{\omega_{k+}x+\omega_{k-}x_b}-\omega_{k+} e^{\omega_{k-}x+\omega_{k+}x_b}}
{\omega_{k+} e^{\omega_{k+}x_b}-\omega_{k-} e^{\omega_{k-}x_b}} \ , \\ \nonumber
\omega_{k\pm}&=\sqrt{\Omega}\pm\sqrt{\Omega-1+k^2}\ .
\end{align}

For $\Omega>1$ the limit $x_b\to \infty$ can easily be done and we get the following expression for the average volume of type $y_r$
\begin{align} \label{eq:Vav1}
\langle V(y_r)\rangle &=\int_{-\infty}^{\infty} \frac{dk}{2\pi}e^{i\,k\,\Delta y+\omega_{k-}x}=\frac{\sqrt{\Omega-1}}{\pi}\frac{x}{\sqrt{x^2+y^2}}e^{\sqrt{\Omega}\,x} 
K_1 [\sqrt{(\Omega-1)(x^2+y^2)}]\ ,
\end{align}
where $K_1[x]\equiv\frac12\int_{0}^\infty dt e^{-\frac{x}{2}\left(t+\frac{1}{t}\right)}$ is the modified Bessel function.

It is  easy to check that the average of the total volume
\begin{align}
\langle V\rangle=\int_{-\infty}^{\infty}dy_r \langle V( y_r)\rangle=\int_{-\infty}^{\infty}dy_r\int_{-\infty}^{\infty} \frac{dk}{2\pi}e^{ik\Delta y+\omega_{k-}x}=e^{\omega_{0-}x}=e^{3N_c\frac{2}{1+\sqrt{1-1/\Omega}}}\,
\end{align}
coincides with the one-field case, eq.~(\ref{Vav1}). From this we see that the phase transition to eternal inflation happens at 
\be
\Omega=1\ .
\ee
This result is not surprising given the shift symmetry in $y$. However from eq.~(\ref{eq:Vav1}) we also have the information on the shape of the volume distribution on the reheating surface. As expected the distribution is peaked around $y_r=y$, which is the classical exit point. 
Let us look at the shape in several limits (see also fig.~\ref{fig:waterfall}).

We can expand the Bessel function to obtain
\be
\langle V(y_r)\rangle\quad\rightarrow\quad e^{-\frac{\sqrt{\Omega-1}}{x}\frac{\Delta y^2}{1+\sqrt{1+\Delta y^2/x^2}}} \ ,\qquad {\rm for}\quad (\Omega-1)(x^2+\Delta y^2)\gg 1\,,
\ee
which is a Gaussian in $\Delta y$ for $\Delta y\lesssim x$ (when the $y$-distance of the starting point from the reheating point is smaller than the classical trajectory). The tail of the Gaussian turns into an exponential for $\Delta y\gtrsim x$ while the width of the Gaussian is always smaller than~$x$.

Instead, very near to the transition to eternal inflation, $\Omega\rightarrow 1$, the distribution approaches a Lorentzian for $\Delta y$ small enough:
\be
\langle V(y_r)\rangle\quad\rightarrow\quad \frac{x}{x^2+\Delta y^2} \ ,\qquad {\rm for}\quad  (\Omega-1)(x^2+\Delta y^2)\ll1 
\ee
while the tail becomes again exponential $e^{-(\sqrt{\Omega-1}\;\Delta y)}$ for $\Delta y\gtrsim (\Omega-1)^{-1/2}$.

The bound on the volume is in this case trivially satisfied. The reheating entropy is the same for every reheating point ($S(y)=S$), and for the volume associated to each reheating region ${\cal I}$ we have
\be
\int_{\cal I} dy_r\; \langle V(y_r)\rangle \leq \langle V\rangle=e^{3N_c\frac{2}{1+\sqrt{1-1/\Omega}}} < e^{-S/2}\ .
\ee


\subsection{Second example: tilted waterfall}


We now consider a generalization of the former case. We imagine that the reheating surface is defined at $x=0 $ as in the former case. However, now the gradient of the inflaton potential is not orthogonal to the reheating surface: $\bm{\sqrt\Omega}\,=(\sqrt\Omega_x,\sqrt\Omega_y)$. Consequently the value of the Hubble radius at each point of the reheating surface can be different, this fact will allow us to test the generalized bound in (\ref{eq:average-bound}) in a non-trivial way. 

In this case eq.~(\ref{eq:master-local}) becomes
\be
\psi_{xx}+\psi_{yy}-2\left[\sqrt{\Omega}_x \psi_{x}+\sqrt{\Omega}_y \psi_{y}\right]+\psi=0\ ,
\ee
with the same boundary conditions\footnote{Actually since in this case there is a non vanishing slope also in the $y$ direction, the barrier will be naturally at angle, normal to the vector $\bm{\sqrt{\Omega}}$, if, for example, corresponds to the region where $H=$const. In particular it will always intersect the reheating surface in one point, as $H$ will be Planckian somewhere
on ${\cal R}$. For this analysis we are interested only in the region far from the barrier, where boundary effects can be neglected. So we took the barrier to be parallel to the
reheating surface as in the previous example and then took the limit $\tau_b\to \infty$. For our purposes this is a safe approximation because, 
as explained in \cite{Dubovsky:2008rf}, finite boundary effects are expected to make the actual bound even stronger since they cut-off inflaton trajectories getting arbitrary far from the reheating surface.}. 
The solution for the average volume is similar to the previous case:
\begin{align} \label{eq:Vav2}
\langle V(y_r)\rangle &=\int_{-\infty}^{\infty} \frac{dk}{2\pi}e^{ik \,\Delta y
+\sqrt{\Omega}_x x+\sqrt{\Omega}_y \Delta y - \sqrt{\Omega-1+k^2}x} \\
	&=\frac{\sqrt{\Omega-1}}{\pi}\frac{x}{\sqrt{x^2+\Delta y^2}}e^{\sqrt{\Omega}_x x+\sqrt{\Omega}_y \Delta y} 
K_1 [\sqrt{(\Omega-1)(x^2+\Delta y^2)}]\ ,
\end{align}
where we defined 
\be
\Omega\equiv\left(\bm{\sqrt\Omega}\right)^2=\Omega_x+\Omega_y \,.
\ee
From the solution above, we notice that this time the phase transition happens when $\Omega_x$ becomes smaller than one, i.e. before $\Omega$ reaches one. Indeed if we consider the total average volume:
\begin{align}
\langle V\rangle=\int_{-\infty}^{\infty}dy_r \langle V(y)\rangle=\int_{-\infty}^{\infty}dy_r\int_{-\infty}^{\infty} 
\frac{dk}{2\pi}e^{ik\,\Delta y+\sqrt{\Omega}_x x+\sqrt{\Omega}_y \Delta y - \sqrt{\Omega-1+k^2}x}=e^{(\sqrt{\Omega}_x-\sqrt{\Omega _x-1})x}\ ,
\end{align}
we can see that it is not analytic at $\Omega_x=1$, signaling the  onset of the eternal inflation regime. This was indeed expected from the symmetries of the problem---it is only the gradient of the potential normal to the reheating surface $\sqrt{\Omega}_x$ that matters in determining the transition to eternal inflation.
The slope $\sqrt{\Omega}_y$ along the $y$-direction, parallel to the reheating surface obviously plays no role for the phase transition.
Still the information about the reheating point $y_r$ is non-trivial in this model, as for example each different volume $y_r$ is associated to a different Hubble scale or more generally 
to a different kind of vacuum. If we look at the average value of the reheating point defined as
\be
\langle  y_r \rangle=\frac{\int_{-\infty}^{\infty}d y_r\; y_r\,\langle V(y_r)\rangle}{\int_{-\infty}^{\infty}d y_r\;\langle V(y_r)\rangle}=y-x\, \sqrt{\frac{\Omega_y}{\Omega_x-1}}\ ,
\ee
we can see that for $\Omega_x\gg 1$, $\langle y_r\rangle$ coincides with the exit point of the classical trajectory $y_r^{\rm cl}=y-x\sqrt{\Omega_y/\Omega_x}$. 
In this limit, $\langle V(y_r)\rangle$ is a Gaussian sharply peaked around $y_{r}^{\rm cl}$, as in the previous example. 
Instead as $\Omega_x\rightarrow 1$, $\langle y_r\rangle\rightarrow -\infty$, and $\langle V(y_r)\rangle$ broadens as roughly shown in Fig.~\ref{fig:waterfall}. 
This result is qualitatively expected: the closer we get to the phase transition,
the longer the stochastic trajectories become, and the more they drift downhill (because of the $\sqrt{\Omega}_y$ tilt). This explains why the singularity in 
$\langle V(y_r)\rangle$ only appears at $\Omega=1$. Indeed even when $\Omega_x<1$, $\langle V(y_r)\rangle$ can be finite for any $y_r$, the infinity of $\langle V\rangle$ 
being due to trajectories exiting at $y_r\to- \infty$. When the whole $\Omega=1$, then also $\langle V(y_r)\rangle$ has to start diverging because trajectories
can start to go up-hill also in the $y$ direction and produce infinite volume at finite $y_r$. Graphically the phase diagram looks like as follows:
\begin{center}
\includegraphics[height=0.36\textwidth]{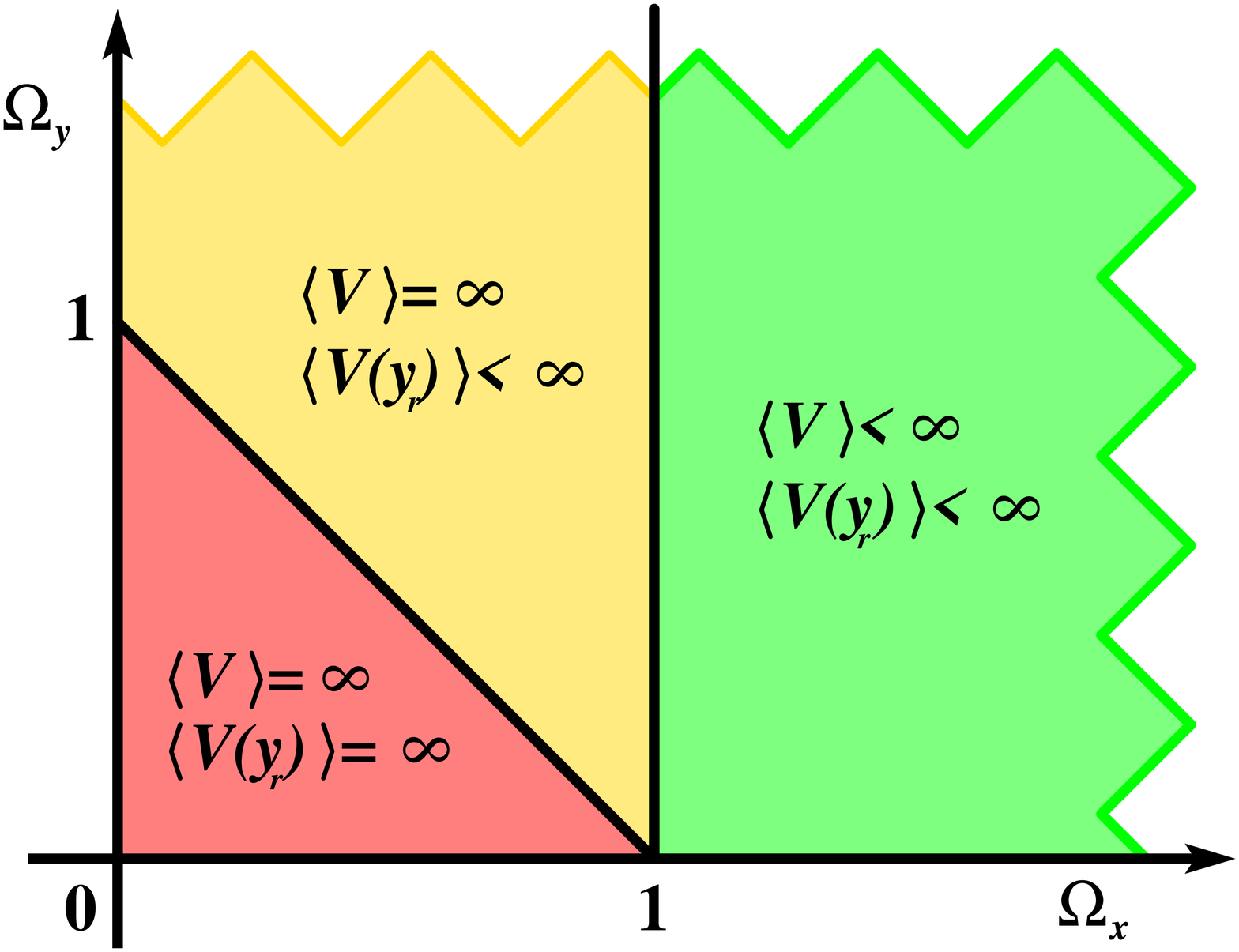}
\end{center}
Notice that for $\Omega_x<1$ but $\Omega>1$ (the yellow region above) we have eternal inflation only because the total volume
diverges, while $\langle V(y_r)\rangle$ stays finite for every finite $y_r$. In realistic models we would expect also the range of $y$ to be finite
(as the shift symmetry in $y$ is broken by $\Omega_y$). In this case the divergence in $\langle V \rangle$ disappears being due to the non integrability
of $\langle V(y_r)\rangle$. The yellow region turns into a non-eternal slow-roll inflation region and the phase transition is expected to happen now at $\Omega=1$.

We are now ready to formulate and test the bound in this multi field case.
Eq.~(\ref{eq:average-bound}) must be checked for any interval ${\cal I}$ on the reheating surface, however since the entropy
in this example is monotonically decreasing with $y_r$ it will be enough to consider just the intervals ${\cal I}=[y_r^0,\infty)$ for every $y_r^0$.
Notice that the total volume is bounded by
\begin{equation}
\langle V\rangle=e^{(\sqrt{\Omega}_x-\sqrt{\Omega_x-1})x}=e^{6N_c \frac{1}{1+\sqrt{1-\frac{1}{\Omega_x}}}}<e^{S_c/2}
\end{equation}
where $S_c$ is the entropy as calculated at the value of the Hubble constant corresponding to the classical exit point $y_r^{\rm cl}$
of the reheating surface and we used the relation $x=6\sqrt{\Omega}_xN_c$. It follows that the bound is trivially satisfied on any interval ${\cal I}=[y_r^0,\infty)$ with $y_r^0<y_c$ since
\begin{equation}
\int_{y_r^0}^{\infty} dy_r\,\langle V(y_r)\rangle < \langle V\rangle < e^{S_c/2} <e^{S(y_r^0)/2}\,.
\end{equation}
We just need to show now that the bound is not violated when $y_r^0>y_r^{\rm cl}$. 

For $(\Omega-1)(x^2+\Delta y^2)\gg 1$, $\langle V(y_r)\rangle$ behaves like
\begin{equation}
\langle V(y_r)\rangle \sim e^{\sqrt{\Omega}_x x+\sqrt{\Omega}_y \Delta y-\sqrt{(\Omega-1)(x^2+\Delta y^2)}}\,,
\end{equation}
which is a falling exponential for $y_r>y_r^{\rm cl}$ (since $\langle V(y_r)\rangle$ is peaked around $\langle y_r\rangle<y_r^{\rm cl}$).
Therefore the integral in the l.h.s. of  eq.~(\ref{eq:average-bound}) is dominated by the value of the integrand at $y_r^0$.
We thus need to check whether
\begin{equation}\label{ineqexp}
\langle V(y_r^0) \rangle <e^{\frac{S(y_r^0)}{2}} \,,
\end{equation}
up to pre-exponential factors.
At leading order in the slow roll approximation we also have that $S(y_r^0)=S_c-2\sqrt{\Omega_y} (y_r^0-y_r^{\rm cl})$ and looking only
at the exponents of (\ref{ineqexp}) we have
\begin{equation}
\sqrt{\Omega_x}x+\sqrt{\Omega_y}\Delta y-\sqrt{(\Omega-1)(x^2+\Delta y^2)}<\frac{S_c}{2}-\sqrt{\Omega_y}(y_r^0-y_r^{\rm cl})\,.
\end{equation}
Using the fact that $x=6\sqrt{\Omega_x}N_c$ and defining $\delta y=y_r^0-y_r^{\rm cl}$, after a bit of algebra we get
\begin{equation} \label{ineqexp2}
6\sqrt{\Omega}N_c\left (\sqrt{\Omega-1}\sqrt{1+\frac{\Omega_x}{\Omega}\frac{\delta y}{x}
\left(\frac{\delta y}{x}-2\sqrt{\frac{\Omega_y}{\Omega_x}} \right) }+\sqrt{\Omega}\left(\frac{S_c}{12 \Omega N_c }-1\right) \right)>0\,.
\end{equation}
This is the sum of two terms. The first is positive and the second is also positive since $S_c>\Delta S=S_c-S_{start}=12  \Omega N_c$.

Of course, the bound continues to hold even when $(\Omega-1)(x^2+\Delta y^2)\gg 1$ is not satisfied---right before entering 
the phase of eternal inflation---because the distribution of volumes move further towards large negative values of $y_r$ where
the bound is trivially satisfied because the volume produced in ${\cal I}$ is smaller.

Notice also that the l.h.s. of (\ref{ineqexp2}) is minimized at $\delta y=\Delta y_c$ (corresponding to $y_r^0=y$), where we get
\begin{equation}
6N_c\left[\sqrt{\Omega_x}\sqrt{\Omega-1}+\Omega\left(\frac{S_c}{12\Omega N_c}-1\right)\right]>0\,,
\end{equation}
which can be saturated only when $S_c=\Delta S$ and $\Omega=1$, which implies $\Omega_y=0$, i.e. the untilted potential of the previous section,
corresponding in practice to single-field inflation.

This shows that tilting the reheating surface with respect to the gradient in field space actually makes the bound stronger.
It also suggests that only single field slow-roll inflation manages to actually saturate the bound, any other multifield models
which move further away from the symmetric setup of the single field case, seem to make the bound stronger, by producing less volume.






\section*{Acknowledgments}

We would like to thank Nima Arkani-Hamed for  encouragement and Juan Maldacena, Steve Shenker, Eva Silverstein 
and Neil Turok for discussions.
GV was partially supported by ERC grant BSMOXFORD no. 228169. The work of SD is supported in part by the NSF grant PHY-1068438.
SD also thanks the hospitality of the Aspen Center for Physics
(under the NSF grant No. 1066293) and the Stanford Institute of Theoretical Physics 
where part of this work has been done.

\appendix

\section{Some explicit computations}
\label{app:one}

In this appendix we report some of the calculations omitted in section~\ref{sec:hd}.

\begin{figure}[t]
\begin{center}
\includegraphics[height=0.36\textwidth]{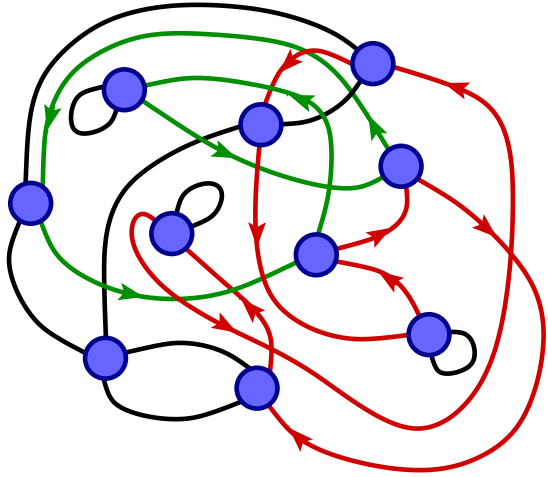} 
\caption{\label{fig:net} \small \it Each node corresponds to the insertion of a Riemann tensor,
each link to the contraction of a pair of indices. Closed orientable loops can be identified such that
all nodes will end up having an equal number of incoming and outgoing lines. Hence any scalar
contraction of Riemann tensors can be written just in terms of Riemann tensors with two upper and two
lower indices contracted without using the metric.}
\end{center}
\end{figure}

First we show that any given Lagrangian $\cal L$, generic function of the metric $g_{\mu\nu}$, the Riemann tensor 
$R_{\mu\nu\rho\sigma}$ and scalar quantities $\phi$, with no explicit covariant derivatives, can be rewritten
just in terms of the Riemann tensors with two covariant and two contravariant 
indices ($R^{\mu\nu}{}_{\rho\sigma}$ and $R^{\mu}{}_{\nu}{}^{\rho}{}_{\sigma}$) without any explicit dependence
on the metric $g_{\mu\nu}$, i.e.
\begin{equation} \label{eq:statement}
{\cal L}(g_{\mu\nu},\, R_{\mu\nu\rho\sigma}, ...,\phi)=\widetilde {\cal L}(R^{\mu\nu}{}_{\rho\sigma}, \, R^{\mu}{}_{\nu}{}^{\rho}{}_{\sigma}, ...,\phi)\,.
\end{equation}
Any term in ${\cal L}$ can be viewed as a network of nodes and lines, each node corresponding to the insertion
of a Riemann tensor, each line to the contraction of two indices with a metric. Each node has thus four
lines attached to it. The statement (\ref{eq:statement}) corresponds to showing that the lines of each network
can be oriented such that each node has two incoming and two outgoing lines (see fig.~\ref{fig:net}).

Consider now a node of the network and let start following a line departing from it. Since
every node has an even number of lines attached, when the line arrives to a node there will be another 
line which has not been used yet, which can be followed, until eventually the path will close 
arriving back to the initial node. At this point we have a closed path which we can orient. Each
node in this path will have an equal number of incoming and outgoing lines and an even number
of unused lines. We can thus repeat the construction by starting with another unused line,
forming another closed loop and orienting it. When we will have used all the available links we will
have oriented all lines such that an equal number of incoming and outgoing line will pass through
each node. This proves eq.~(\ref{eq:statement}).

We prove now that the equations of motion (\ref{eq:eegen}) can be rewritten as in eq.~(\ref{eq:eegen2}).
First we use that
\begin{align}
\widetilde {\cal L}(R^{\mu\nu}{}_{\rho\sigma}, \, R^{\mu}{}_{\nu}{}^{\rho}{}_{\sigma}, ...,\phi)&=
\widetilde {\cal L}(R_{\alpha\beta\rho\sigma} g^{\alpha\mu} g^{\beta\nu}, \, R_{\alpha\nu\beta\sigma}g^{\alpha\mu} g^{\beta\rho}, ...,\phi) \nonumber \\
\frac{\delta \widetilde {\cal L}}{\delta R^{\mu\nu}{}_{\rho\sigma}}&=
\frac{\delta_1 \widetilde {\cal L}}{\delta_1 R_{\alpha\beta\rho\sigma}}g_{\alpha\mu} g_{\beta\nu}\nonumber \\
\frac{\delta \widetilde {\cal L}}{\delta R^{\mu}{}_{\nu}{}^{\rho}{}_{\sigma}}&=
\frac{\delta_2 \widetilde {\cal L}}{\delta_2 R_{\alpha\nu\beta\sigma}}g_{\alpha\mu} g_{\beta\rho}\nonumber
\end{align}
where $\delta_{n}/\delta_{n}R_{\mu\nu\rho\sigma}$ means that we only differentiate with respect to the $R_{\mu\nu\rho\sigma}$ appearing in the $n$-th argument.
We thus have that 
\begin{align}
\frac{\delta \widetilde{\cal L}}{\delta R^{\alpha\beta}{}_{\rho\sigma}}\delta R^{\alpha\beta}{}_{\rho\sigma}+
\frac{\delta \widetilde{\cal L}}{\delta R^{\alpha}{}_{\beta}{}^{\rho}{}_{\sigma}}\delta R^{\alpha}{}_{\beta}{}^{\rho}{}_{\sigma}
&=\frac{\delta_1 \widetilde {\cal L}}{\delta_1 R_{\alpha\beta\rho\sigma}}g_{\alpha\mu} g_{\beta\nu}\delta R^{\mu\nu}{}_{\rho\sigma}+\frac{\delta_2 \widetilde {\cal L}}{\delta_2 R_{\alpha\nu\beta\sigma}}g_{\alpha\mu} g_{\beta\rho}\delta R^{\mu}{}_{\nu}{}^{\rho}{}_{\sigma}\nonumber \\
&=\left (\frac{\delta_1 \widetilde {\cal L}}{\delta_1 R_{\mu\nu\rho\sigma}}+\frac{\delta_2 \widetilde {\cal L}}{\delta_2 R_{\mu\nu\rho\sigma}}\right) \left ( \delta R_{\mu\nu\rho\sigma}-2 R^{\alpha}{}_{\nu\rho\sigma}\delta g_{\alpha \mu} \right) \nonumber \\
&=\frac{\delta {\cal L}}{\delta R_{\mu\nu\rho\sigma}}\left( 2\nabla_{\mu}\nabla_{\sigma}\delta g_{\nu\rho}-R^{\alpha}{}_{\nu\rho\sigma}\delta g_{\alpha \mu} \right)
\end{align}
where in the last step we used the identity (\ref{eq:waldid}) and the fact that 
\begin{equation}
\frac{\delta_1 \widetilde {\cal L}}{\delta_1 R_{\mu\nu\rho\sigma}}+
\frac{\delta_2 \widetilde {\cal L}}{\delta_2 R_{\mu\nu\rho\sigma}}=
\frac{\delta {\cal L}}{\delta R_{\mu\nu\rho\sigma}}\,.
\end{equation}
Hence, we finally have
\begin{equation}
\frac{\delta \widetilde{\cal L}}{\delta R^{\alpha\beta}{}_{\rho\sigma}}\frac{\delta R^{\alpha\beta}{}_{\rho\sigma}}{\delta g_{\mu\nu}}+
\frac{\delta \widetilde{\cal L}}{\delta R^{\alpha}{}_{\beta}{}^{\rho}{}_{\sigma}}\frac{\delta R^{\alpha}{}_{\beta}{}^{\rho}{}_{\sigma}}{\delta g_{\mu\nu}}
=2\nabla_{\rho}\nabla_{\sigma}\frac{\delta {\cal L}}{\delta R_{\rho(\mu\nu)\sigma}} 
-R^{(\mu}{}_{\alpha\beta\gamma}\frac{\delta {\cal L}}{\delta R_{\nu)\alpha\beta\gamma}}\,,
\end{equation}
which can be used to derive eq.~(\ref{eq:eegen2}) from eq.~(\ref{eq:eegen}).

The last missing step to explicitly show the universality of the bound with respect to higher derivative terms,
is the relation between the variation of the de~Sitter entropy and the metric equations of motion,
in particular how to get eq.~(\ref{eq:dsdtinter}) from the variation of  eq.~(\ref{eq:waldS}).
We start with the Wald formula for the entropy of $\Sigma(t)$, which reads
\begin{equation}
S(t)=-\frac{4\pi}{\kappa}\int_{\Sigma(t)} d\Sigma_{\mu\nu} \left(
\frac{\delta {\cal L}}{\delta R_{\mu\nu\rho\sigma}}\nabla_\rho \xi_{\sigma}
-2\nabla_{\rho}\frac{\delta {\cal L}}{\delta R_{\mu\nu\rho\sigma}} \xi_{\sigma}\right)\,,
\end{equation}
where $\kappa$ is the surface gravity on the stretched horizon.
Using Stokes theorem it follows that 
\begin{align} \label{eq:findsdt}
\frac{dS}{dt}&=\frac{4\pi}{\kappa}\int_{\Upsilon}\frac{d\tau}{dt} dA\, n_\mu \nabla_{\nu}
\left(\frac{\delta {\cal L}}{\delta R_{\mu\nu\rho\sigma}}\nabla_\rho \xi_{\sigma}
-2\nabla_{\rho}\frac{\delta {\cal L}}{\delta R_{\mu\nu\rho\sigma}} \xi_{\sigma}\right) \nonumber\\
&=\frac{4\pi}{\kappa}\int_{\Upsilon}dA\, {\tilde n}_\mu\left[ \nabla_{\nu}
\left(\frac{\delta {\cal L}}{\delta R_{\mu\nu\rho\sigma}}-2\frac{\delta {\cal L}}{\delta R_{\mu\rho\nu\sigma}}
\right) \nabla_\rho \xi_{\sigma}+\frac{\delta {\cal L}}{\delta R_{\mu\nu\rho\sigma}} \nabla_\nu \nabla_\rho\xi_{\sigma}
-2\nabla_\nu\nabla_{\rho}\frac{\delta {\cal L}}{\delta R_{\mu\nu\rho\sigma}} \xi_{\sigma}\right] \nonumber\\
&=\frac{4\pi}{\kappa}\int_{\Upsilon}dA\, {\tilde n}_\mu\xi_{\sigma}\left(\frac{\delta {\cal L}}{\delta R_{\mu\nu\rho\tau}} R^{\sigma}{}_{\nu\rho\tau}
-2\nabla_\nu\nabla_{\rho}\frac{\delta {\cal L}}{\delta R_{\mu\nu\rho\sigma}} \right) \nonumber\\
&\stackrel{\alpha\to 0}{\longrightarrow} \frac{4\pi}{\kappa}\int_{\cal H} dA\, {\xi}_\mu \xi_{\nu}
\left(\frac{\delta {\cal L}}{\delta R_{\mu\sigma\rho\tau}} R^{\nu}{}_{\sigma\rho\tau}
-2\nabla_\sigma\nabla_{\rho}\frac{\delta {\cal L}}{\delta R_{\sigma\mu\nu\rho}} \right)\,, 
\end{align}
where in the first step we defined $\tilde n^{\mu}\equiv\frac{d\tau}{dt}n^\mu=(\alpha,-H r/\alpha,0,0)$,
which is normalized like the Killing vector (${\tilde n}^\mu {\tilde n}_\mu=-\xi^\mu \xi_\mu$),
in the second step we used the cyclic properties of the Riemann tensor, in the third the fact that
\begin{equation}
\nabla_{(\mu}\xi_{\nu)}=0\,,\qquad \nabla_{\mu}\nabla_{\nu}\xi_{\rho}=R^{\sigma}{}_{\mu\nu\rho}\xi_{\sigma}\,,
\end{equation}
and in the last we performed the horizon limit $\alpha\to1$.
Eq.~(\ref{eq:findsdt}) matches eq.~(\ref{eq:dsdtinter}) and this terminates our proof.


\end{document}